\documentclass{ws-p8-50x6-00}

\def\thm{\theta_\mu}
\def\tha{\theta_A}

\def\m12{m_{1\!/2}}

\def\msf{m_{\tilde f}}
\def\mchi{m_{\tilde \chi}}
\def\ohsq{\Omega_{\widetilde\chi}\, h^2}
\def\ch{\widetilde \chi}

\def\gev{{\rm \, Ge\kern-0.125em V}}
\def\ga{\mathrel{\raise.3ex\hbox{$>$\kern-.75em\lower1ex\hbox{$\sim$}}}}
\def\la{\mathrel{\raise.3ex\hbox{$<$\kern-.75em\lower1ex\hbox{$\sim$}}}}
\def\gyr{{\rm \, G\kern-0.125em yr}}
\def\tb{\tan\beta}

\begin{document}

\title{\vspace*{-2cm}\rightline{\rm MADPH-00-1152}\rightline{\rm January, 2000}\vspace{2cm}
CP Violating Phases and the Dark Matter Problem\thanks{\rm Presented at COSMO99: 3rd International Conference on Particle Physics and the Early Universe, Trieste, Italy}}
\author{Toby Falk}
\address{ Department of Physics, University of Wisconsin, Madison,
 WI~53706, USA}

\maketitle

\abstracts{New CP violating phases in the MSSM can affect both the
  abundance and detection of neutralino dark matter.  We discuss
  the effect of including cosmological constraints in the limits on new
  sources of CP violation in the MSSM and the effects of new CP
  violating parameters on dark matter densities and detection.
}

\section{CP Violation in Supersymmetry}
This last year has seen a lot of work on CP violating phases in
supersymmetry--how to constrain the phases, how to measure them, how
to avoid the same constraints, and the extent to which CP violation
can spoil predictions appropriate in the absence of CP violation in
the SUSY parameters.  Today I will discuss some of the cosmological
consequences of CP violating phases in the MSSM, and in particular,
their effect on the abundance and detection of SUSY dark matter.

The Supersymmetric Standard Model contains many new potential sources
of CP violation beyond that of the standard model.  In particular, the
supersymmetric Higgs mixing mass $\mu$, the gaugino masses $M_i$, the
scalar trilinear couplings $A_i$ and the SUSY breaking scalar Higgs
mixing parameter $B\mu$ can all in principle be complex.  However, not
all the phases are physical, and depending on the model, some or most
can be removed by field redefinitions. The remaining sources for CP
violation are experimentally constrained, primarily due to their
contributions to the Electric Dipole Moments (EDMs) of the electron
and neutron, and in particular, the EDM of the mercury atom
$^{199}$Hg.

\section{mSUGRA Constraints}

In minimal Supergravity (mSUGRA), the large number of relations
between the SUSY parameters reduces the set of CP violating phases to
just two: $\theta_\mu$, associated with the Higgs mixing mass $\mu$,
and $\theta_A$, a common trilinear parameter phase.  These phases then
appear in the low energy Lagrangian in the neutralino and chargino
mass matrices (in the case of $\theta_\mu$) and in the left-right
sfermion mixing terms (both $\theta_\mu$ and $\theta_A$).  The new
sources for CP violation then contribute to the EDMs of standard model
fermions, and the tight experimental constraints on the EDMs of the
electron, neutron and mercury atom place severe limits on the sizes of
$\theta_\mu$ and $\theta_A$ \cite{ko,fopr}.

The EDMs generated by $\theta_\mu$ and $\theta_A$ are sufficiently
small if either 1)~the phases are very small ($\la 10^{-2}$), or 
2)~the SUSY masses are very large (${\cal O}$~(a few TeV)), or 3)~There
are large cancellations between different contributions to the EDMs.
In mSUGRA, option 2) is forbidden by the relic density constraints, as 
we'll show next.  Condition 3), large cancellations, does naturally
occur in mSUGRA models over significant regions of parameter space, including
in the body of the cosmologically allowed region with $\m12={\cal 
  O} (100-400\;{\rm GeV})$.  These cancellations relax the constraints
on the phases, but the limit on $\theta_\mu$ remains small,
$\theta_\mu\la\pi/10$.

To see why option 2) is cosmologically forbidden, recall that the SUSY 
phases contribute to the electron EDM, for example, via processes of
the following type:

\begin{picture}(1,65)
\hspace*{.75in}
    \epsfig{file=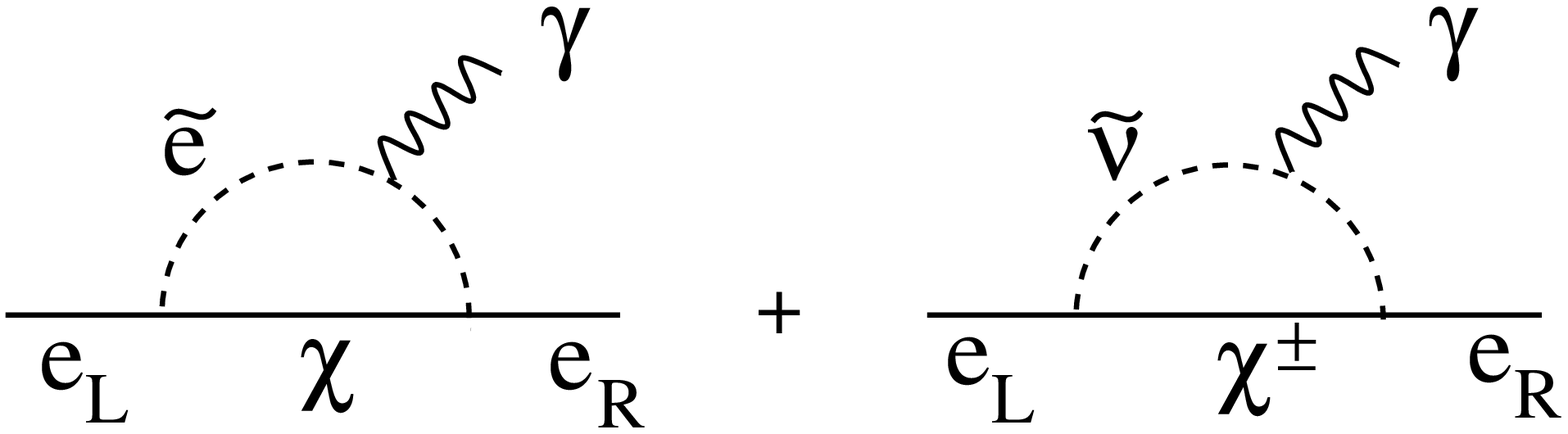, height=0.8in}     
    \label{fig:diags}
\end{picture}

\begin{figure}[h,t]
  \begin{center}
  \epsfig{file=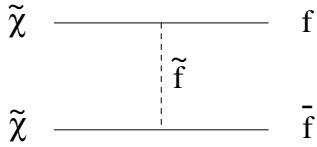, height=0.7in}     
    \caption{Sfermion exchange dominates $\chi$'s annihilation.}
    \label{fig:ann}
  \end{center}
\end{figure}

\begin{figure}[h]
\begin{center}
\begin{minipage}[t]{6in}
\begin{minipage}[t]{2.6in}
\hspace*{-0.1in}
\epsfxsize=2.4in 
\epsfbox{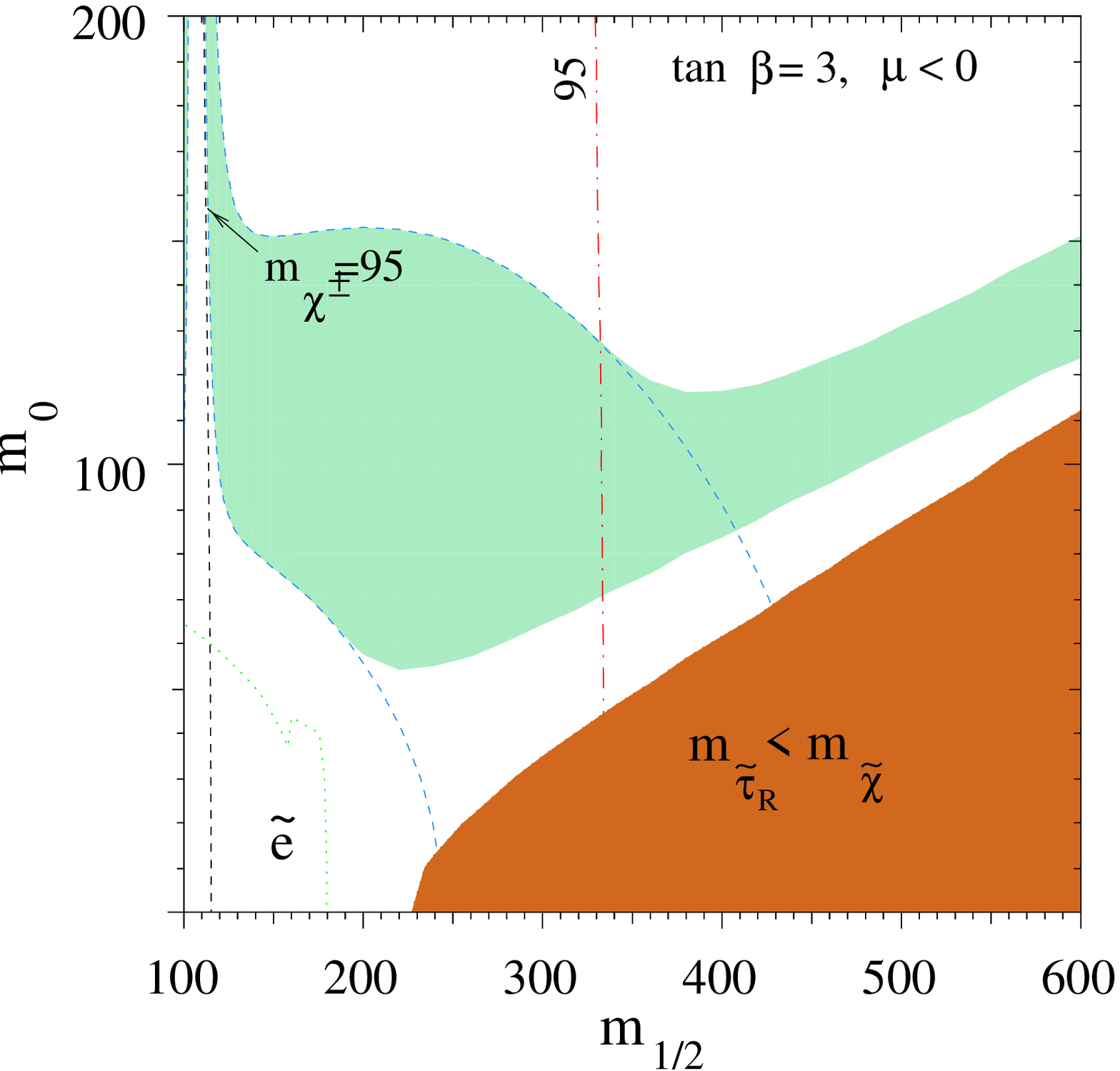} 
\end{minipage}
\hspace*{-0.3in}
\raisebox{-.1cm}{
\begin{minipage}[t]{2.6in}
\epsfxsize=2.4in 
\epsfbox{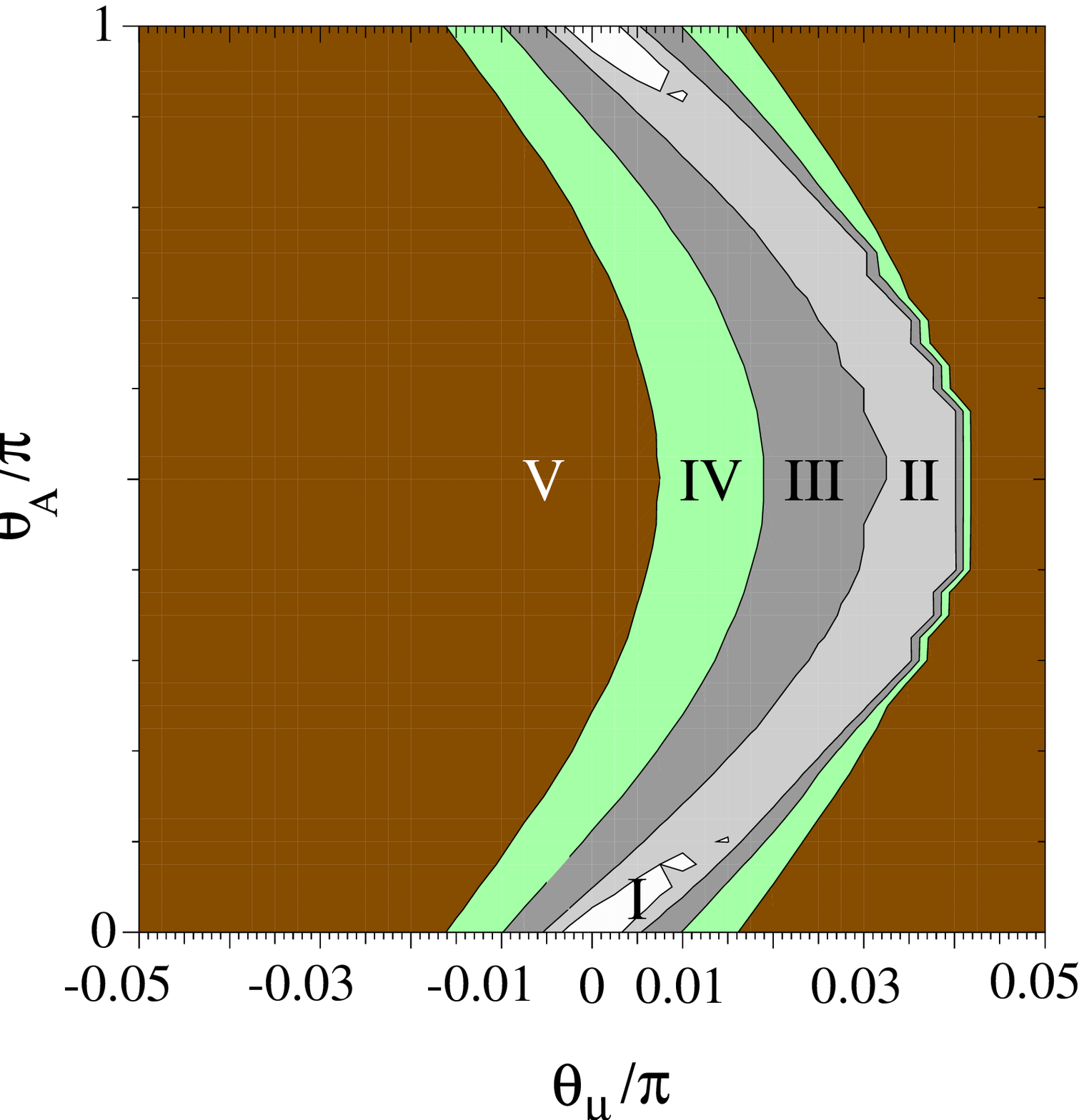} 
\end{minipage}}
\end{minipage}
    \caption{a) The light-shaded area is the cosmologically preferred 
      region with \protect\mbox{$0.1\leq\ohsq\leq 0.3$}.  In the dark
      shaded regions in the bottom right of each panel, the LSP is the
      ${\tilde \tau}_R$, leading to an unacceptable abundance of
      charged dark matter.  Also shown are chargino and Higgs isomass
      contours and an indication of the slepton bound from LEP.
      b) Contours of $\m12^{\rm min}$, the minimum $\m12$ required to
      bring both the electron and Hg EDMs below their respective
      experimental bounds, for $\tan\beta=2, m_0=130\gev$, and
      $A_0=300\gev$. The central light zone labeled ``I'' has
      $\m12^{\rm min}<200\gev$, while the zones labeled ``II'',
      ``III'', and ``IV'' correspond to $200\gev<\m12^{\rm
        min}<300\gev$, $300\gev<\m12^{\rm min}<450\gev$,
      $450\gev<\m12^{\rm min}<600\gev$ and $\m12^{\rm min}>600\gev$,
      respectively.  Zone V is therefore cosmologically excluded.}
    \label{fig:m12min}
\end{center}
\vspace{-0.5cm}
\end{figure}

\noindent{}where selectrons and sneutrinos appear in the loop.  These
contributions diminish as the sfermion masses are increased, but this
also shuts off neutralino annihilation in the early universe, which is
dominated by sfermion exchange as in Fig.~\ref{fig:ann}, and hence
increases the neutralino relic abundance. In
Fig.~\ref{fig:m12min}a we denote\cite{efo} by light shading the region of the
$\m12-m_0$ parameter space with a relic neutralino abundance in the
preferred range $0.1\leq\ohsq\leq0.3$.   The upper
bound on $\ohsq$, coming from a lower limit of 12 Gyr on the age of
the universe, then limits the extent to which one can turn off the
electron EDMs by raising the sfermion masses.  The combination of
cosmological with EDM constraints in the MSSM and mSUGRA is discussed
in detail in \cite{FOS,fko1,fopr}.

To demonstrate the combined limits on $\theta_\mu$ and $\theta_A$ in mSUGRA,
we plot in the $\{\theta_\mu,\theta_A\}$ plane the minimum value of
$\m12$ required to bring the EDMs of both the electron and the mercury
atom ${}^{199}$Hg below their respective experimental constraints
(Fig.~\ref{fig:m12min}b).  These experiments currently provide the
tightest bounds on the SUSY phases\footnote{The extraction of the
  neutron EDM from the SUSY parameter space is plagued by significant
  hadronic uncertainties \cite{fopr}, so that the inclusion of the
  neutron EDM constraint does not improve the limits when the
  uncertainties in the calculated neutron EDM are taken into account}.
Here we've fixed $\tb=2$, $A_0=300$ GeV and $m_0=100$ and scanned
upwards in $\m12$ until the experimental constraints are satisfied.
Due to cancellations, the EDMs are not monotonic in $\m12$; however,
there is still a minimum value of $\m12$ which is allowed.  In the
absence of coannihilations, there is an upper bound on $\m12$ of about
450 GeV (though slightly smaller for this $m_0$); an analogous figure
to Fig.~\ref{fig:m12min}a for $\tb=2$ shows that coannihilations increase
the bound to about 600 GeV.  Comparing with Fig.~\ref{fig:m12min}a, we
see that zone V is cosmologically forbidden, and that the effect of
including coannihilations is to allow zone IV, which was formerly
excluded.

The bowing to the right of the contours in Fig.~\ref{fig:m12min} is a
result of cancellations between different contributions to the EDMs
\cite{FOS}, and we can see that the effect is to relax the upper bound
on $\theta_\mu$ by a factor of a few.  As we increase $A_0$, the
extent of the bowing increases, and larger values of $\theta_\mu$ can
be accessed.  This loophole to larger $\theta_\mu$ is limited by the
diminishing size of the regions in which there are sufficient
cancellations to satisfy the EDM constraints.  In general, the regions
of cancellation for the electron EDM are different than those for the
${}^{199}$Hg EDM, and the two regions do not always overlap.  As
$\theta_\mu$ is increased, the sizes of the regions of sufficient
cancellations decrease; in Fig.~\ref{fig:m12min}, the width in $\m12$
of the combined allowed region near the $\theta_\mu$ upper bound is
40-80 GeV, which on a scale of 200-300 GeV is reasonably broad.
Larger $A_0$ permits larger $\theta_\mu$, but the region of
cancellations shrinks so that a careful adjustment of $\m12$ becomes
required to access the largest $\theta_\mu$.  At the end of the day,
values of $\theta_\mu$ much greater than about $\pi/10$ cannot satisfy
the EDM constraints without significant fine-tuning of the mass
parameters.  At larger values of $\tb$, the upper bound decreases
roughly as $1/\tb$.  See \cite{fopr} for more details on the status of
EDM and cosmological constraints on CP violating phases in mSUGRA.

\section{Large Phases}

Much of the work on SUSY CP violation in the last year has been
inspired by the hope of having large ($\cal O$(1)) phases, and there
have been several suggestions as to how this might be achieved while
satisfying the stringent constraints from the EDMs.  First, the
presence of cancellations between different contributions to the
fermion EDMs (which recently has dubbed the ``cancellation mechanism''
\cite{in98}) has been used to motivate interest in large phases,
although as we have seen in the last section, in mSUGRA, cosmological
considerations limit the extent to which cancellations can free the
phases. If gaugino mass unification is broken, then there are two
additional phases in the model, namely the relative phases between
$M_1$, $M_2$ and $M_3$.  More phases then lead to more opportunities
for cancellations \cite{bgk}.  There may be hints that string theory
can provide (small) regions with large phases in models without
gaugino mass unification \cite{bekl}.

Alternatively, note that the one-loop diagrams contributing to the
fermion EDMs only contain first generation sfermions.  Hence in models
in which the first (or first two) generation sfermions are extremely
heavy \cite{heavy}, the EDMs are suppressed even with $\cal O$(1)
phases.  Of course if the LSP neutralino is gaugino-like, the third
generation sfermions must be quite light in order to satisfy the relic
density constraints.  The phases in these models are still not
completely unconstrained however.  The third generation sfermions can
contribute to the EDMs at two loops \cite{ckp}, and further, phases in
the stop sector (i.e. $\thm,\theta_{A_t}$) enter radiatively into the
Higgs potential\cite{pil} and induce a phase misalignment between the
Higgs vevs, and this can potentially introduce visible effects.  Both
the latter effects are particularly important at large $\tb$.

\section{Neutralino Relic Density}

As originally shown in \cite{FOS}, CP violating phases can have a
large effect on the relic density of the Lightest Supersymmetric
Particle (LSP) in SUSY models.  This occurs because the dominant
annihilation channel for a gaugino-like neutralino
(Fig.~\ref{fig:ann}) exhibits ``p-wave suppression''.  That
is, if one expands the thermally averaged annihilation cross-section
at freeze-out in powers of $(T/\mchi)$,
\begin{equation}
  \label{eq:sigv}
  \langle\sigma_{\ch\ch} v\rangle = a + b (T/\mchi) + {\cal O}(T/\mchi)^2
\end{equation}
the zeroth order term $a$ is suppressed\cite{hg} by $m_f^2$.  This suppresses
the annihilation rate in the early universe, and enhances the $\chi$
relic abundance, by more than an order of magnitude.  However, in the presence
of left-right sfermion mixing and CP violating phases, the zeroth
order term has a piece
\begin{equation}
  \label{eq:acp}
  a \approx {g_1^4\over 32\pi}{\mchi^2\over (\msf^2+\mchi^2-m_f^2)^2}
Y_L^2 Y_R^2 \sin^2  2\theta_f \sin^2\gamma_f + {\cal O}(m_f \mchi)
\end{equation}
where $\theta_f$ is the sfermion mixing angle and $\gamma_f =
Arg(A_f^* + \mu\tan\beta)$.  For significant $\theta_f$ and
$\gamma_f$, this results in a dramatic increase in the annihilation
rate and decrease in the $\chi$ relic density, and it weakens the
cosmological upper bound on $\mchi$ (in the absence of
coannihilations) from $\sim 250$ GeV to $\sim 650$ GeV \cite{FOS}.  In
mSUGRA, however, neutralino annihilation is primarily through sleptons
into lepton pairs, and the mixing angles $2\theta_f$ are typically
very small. The effect of CP phases on the $\chi$ annihilation in
mSUGRA is therefore negligible\cite{fko1}.  This is also generally
true in models where the sfermion mixing angles are small; e.g. for a
string inspired model, see \cite{ks99}.

Phases also potentially affect masses and couplings of the neutralino
and charginos, stop and Higgs particles, as well as
providing mixing between the scalar and pseudoscalar Higgs\cite{pw99}, 
and these can effect the neutralino relic density, particularly for a
Higgsino-type neutralino.

\section{Neutralino Direct Detection}

Phases can also affect the direct detection of relic neutralinos.
In direct detection schemes, relic neutralinos elastic scatter off
of nuclei in a target material, depositing a detectable amount of energy.
The low-energy effective four-Fermi Lagrangian for neutralino-quark
interactions takes the form
\begin{eqnarray}
  \label{eq:leff}
  {\cal L} &=& \bar{\chi} \gamma^\mu \gamma^5 \chi \bar{q_{i}} 
\gamma_{\mu} (\alpha_{1i} + \alpha_{2i}  \gamma^{5}) q_{i} +
\alpha_{3i}  \bar{\chi} \chi \bar{q_{i}} q_{i} 
+ \alpha_{4i} \bar{\chi} \gamma^{5} \chi \bar{q_{i}} \gamma^{5} q_{i}+\nonumber\\
&&\alpha_{5i} \bar{\chi} \chi \bar{q_{i}} \gamma^{5} q_{i} +
\alpha_{6i} \bar{\chi} \gamma^{5} \chi \bar{q_{i}} q_{i},
\end{eqnarray}
The coefficients $\alpha_2$ and $\alpha_3$ contribute to spin-dependent
and spin-independent neutralino-nucleon scattering, respectively.
The phases enter into the coefficients $\alpha_i$ \cite{ffo,cin}. 

\begin{figure}[h,t]
\begin{center}
\vspace{-0.4cm}
\begin{minipage}[b]{7in}
\epsfxsize=2.4in 
\epsfbox{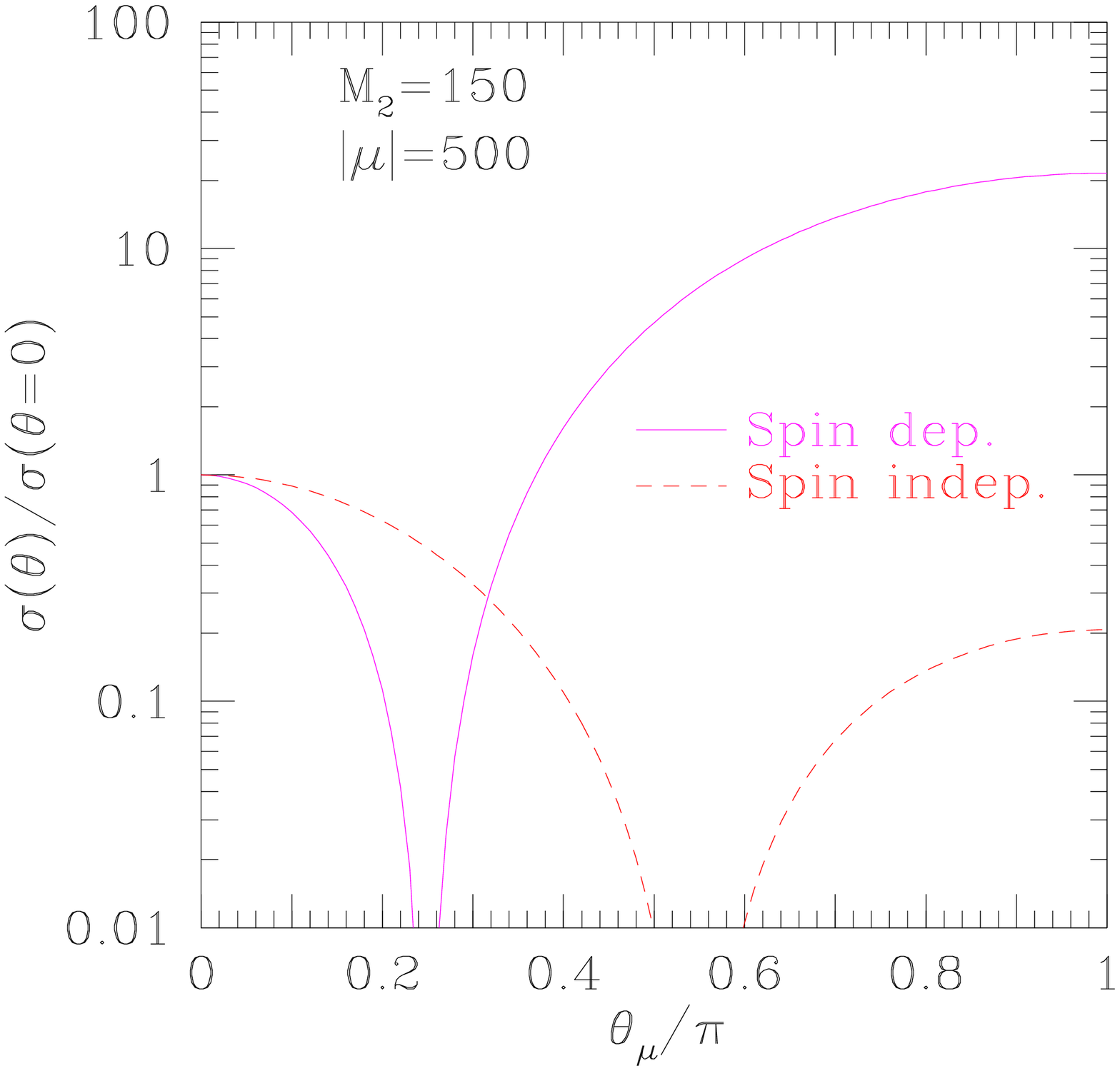}
\hspace{-0.1in}
\epsfxsize=2.4in 
\epsfbox{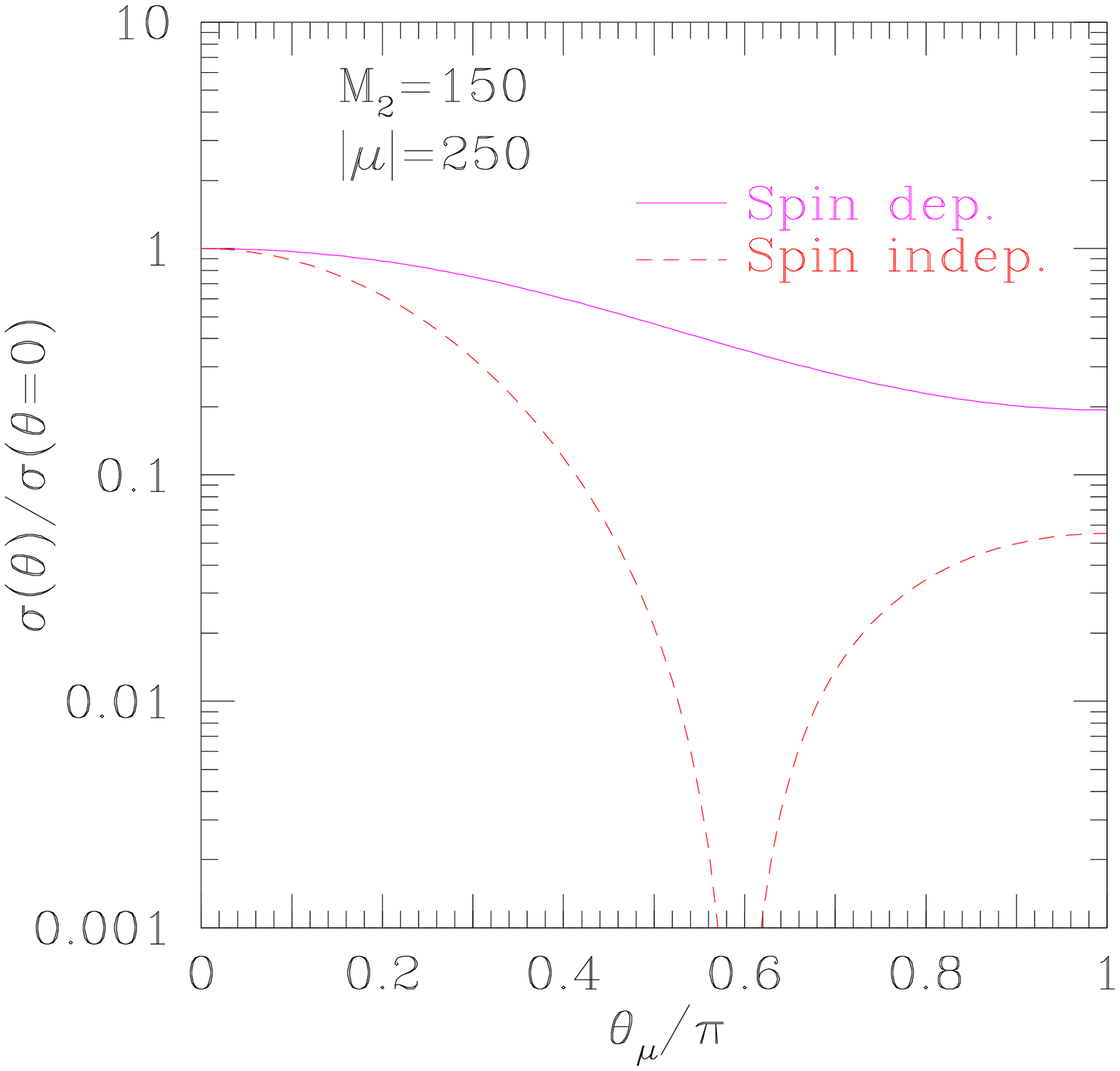}
\end{minipage}
\vspace{-0.8cm}
\caption{Direct detection rates as a function of $\thm$ for
    scattering off of $^{19}$F and b)$^{73}$Ge. Here $\tb=3$ and
    $m_0$=100 GeV. \label{fig:svthm}}
\end{center}
\vspace{-0.4cm}
\end{figure}

Due to cancellations in the scattering rates, phases can produce a
significant effect on the detection rate.  In Fig.~\ref{fig:svthm}, we
show the neutralino-nucleus elastic scattering cross-section as a
function of $\thm$, displaying separately the spin-dependent and
spin-independent contributions, for two target nuclei, $^{19}$F and
$^{73}$Ge.  Dramatic reductions in the spin-independent cross-section
occur near $\thm=0.6\pi$ in both cases, and near $\thm=0.25\pi$ for
the spin-dependent rate for $^{19}$F.  Of course these large values of
$\thm$ are excluded in mSUGRA, as we have seen above, although in more
general models one can tune the other parameters (e.g. the
trilinear parameters $A_i$) in order produce the cancellations
necessary to satisfy the EDM constraints, as we have done in 
Fig.~\ref{fig:svthm}.

\begin{figure}[h,t]
\begin{center}
\begin{minipage}[b]{7in}
\epsfxsize=2.4in 
\epsfbox{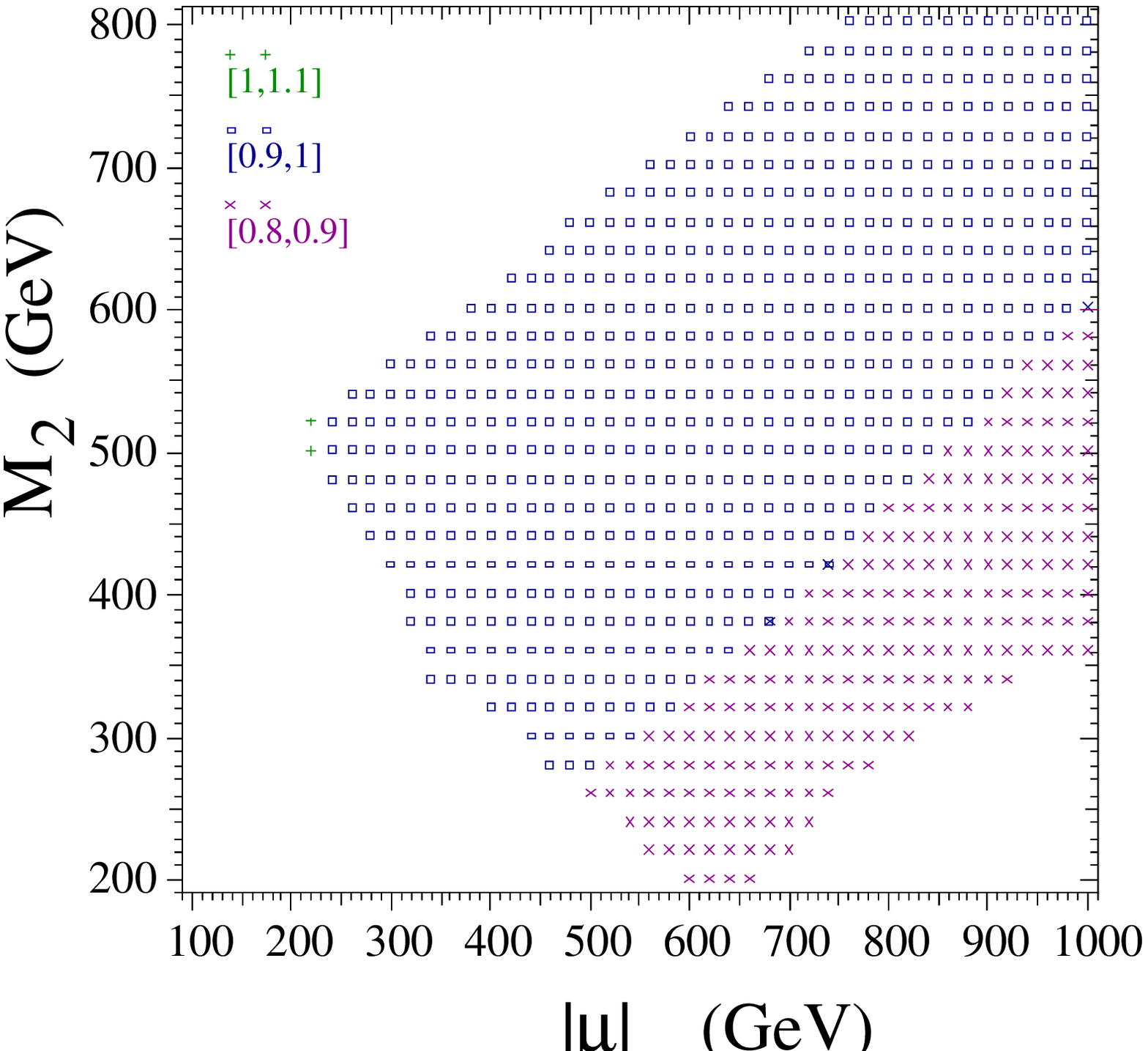}
\epsfxsize=2.52in 
\epsfbox{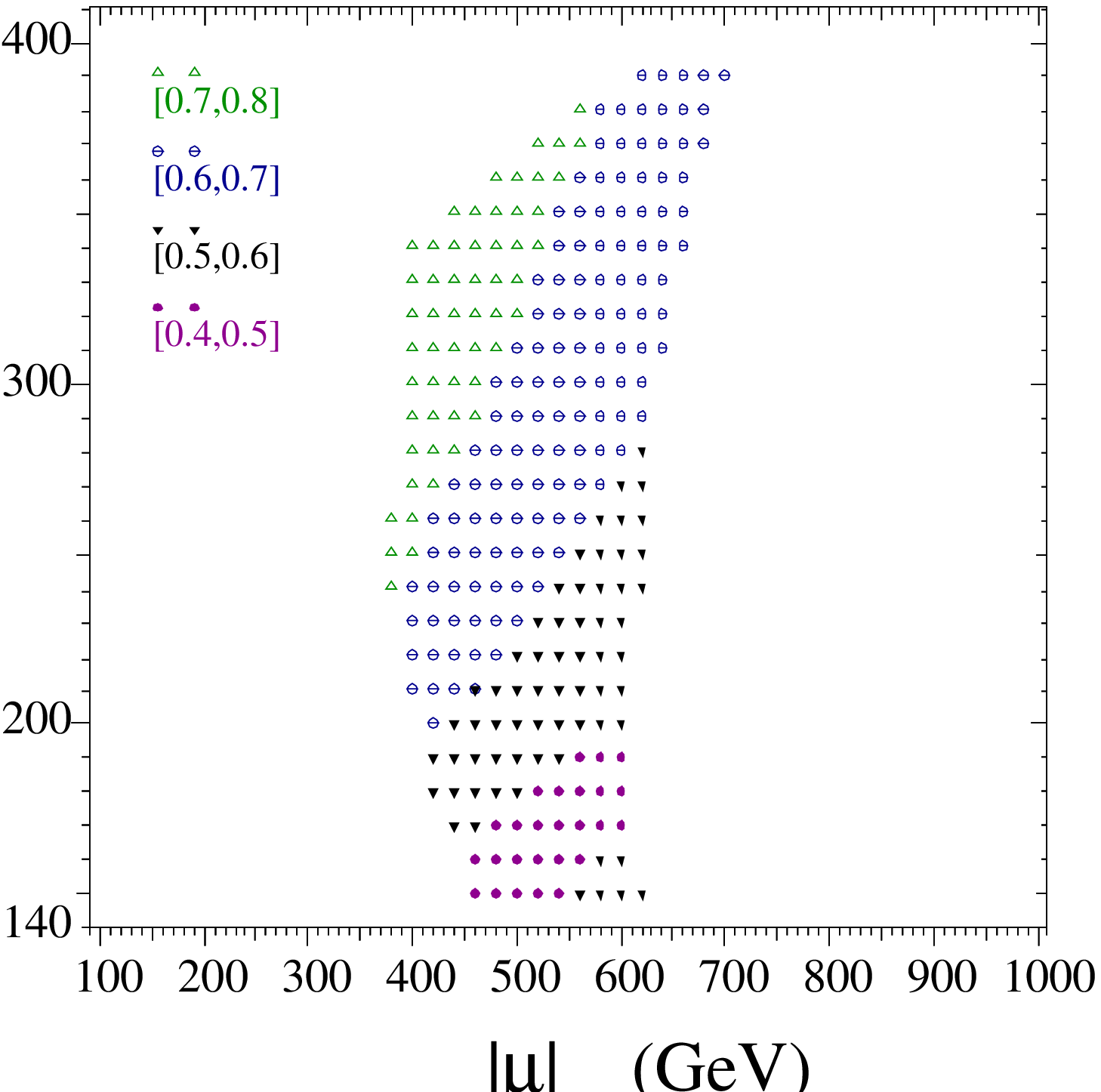}
\end{minipage}
\caption{Ratios of direct detection rates for scattering off of $^{19}$F with and without CP
    violation, for a)$\thm=\pi/8, \tha=3\pi/8$, b)$\thm=\pi/4, \tha=\pi/2$ \label{fig:dd}}
\end{center}
\vspace{-0.7cm}
\end{figure}

In Fig.~\ref{fig:dd} we perform a scan over MSSM parameters
$M_2,\mu,A$ and $m_0$ for fixed phases $\thm$ and $\tha$ and $\tb=3$
and compute the ratio of total scattering cross-sections with and
without phases, for scattering off of $^{19}$F.  We haven't chosen the
phases to lie in the dips of Fig.~\ref{fig:svthm}, so
Fig.~\ref{fig:dd} isn't intended to indicate the maximum possible
effect of the phases.  Rather, we hope display a more typical result
and to demonstrate the variation of the effect of the phases as a
function of the parameters $\mu$ and $M_2$. We see that reductions in
the rate up to \%50 and enhancements by up to \%10 occur over much of
the $M_2-\mu$ parameter space.  The plotted points all satisfy the EDM
constraints.  There is one caveat to bear in mind: because we have
done a scan over parameters, we have found the (small) regions of
parameter space satisfying the EDM bounds for these large values of
the phases.  These regions are not generic, and in fact are
uncomfortably tuned.  Thus we simply take these plots as an existence
proof that CP-violating phases can have a significant effect on the
direct detection of neutralino dark matter.

\section{Summary}

New sources of CP violation are present in the MSSM which are not
present in the standard model.  In mSUGRA, cosmological bounds on
$\m12, m_0$ and $\mchi$ combine with limits on the Electric Dipole
Moments of the electron and $^{199}$Hg to constrain $\thm\la\pi/10$,
while $\tha$ remains essentially unconstrained.  In general models,
phases can affect neutralino annihilation, so that the cosmological
upper bound on $\mchi$ increases from 250 to 650 GeV.  Phases can also 
affect neutralino direct detection rates, typically reducing them by a
factor $\sim 2$, but orders of magnitude in parts of the parameter space.

\section*{Acknowledgments}
The work of T.F.~was supported in part by DOE grant
DE--FG02--95ER--40896, and in part by the University of Wisconsin
Research Committee with funds granted by the Wisconsin Alumni Research
Foundation.


\begin{thebibliography}{99}

 \bibitem{ko}M. Dugan, B. Grinstein and L. Hall, Nucl. Phys. {\bf B255}, 413
(1985); Y. Kizukuri \& N. Oshimo, Phys. Rev. {\bf D45} (1992) 1806;
{\bf D46}(1992) 3025.
\bibitem{fopr} T. Falk, K.A. Olive, M. Pospelov and R. Roiban, hep-ph/9904393.
\bibitem{efo} J. Ellis, T. Falk, and K.A. Olive, Phys. Lett. {\bf B444} (1998)
367.
\bibitem{FOS}  T. Falk, K.A. Olive and M. Srednicki, Phys.Lett. {\bf
B354}  (1995) 99; T. Falk and K.A. Olive, Phys. Lett. {\bf B439} (1998) 71.
\bibitem{fko1}T. Falk and K.A. Olive, Phys. Lett. {\bf 375} (1996) 196.
\bibitem{in98} T.\ Ibrahim, P.\ Nath, Phys. Rev.{\bf D58} (1998) 111301;
  Erratum-ibid.{\bf D60} (1999) 099902. 
\bibitem{bgk} M. Brhlik, G. J. Good and G.L. Kane, Phys. Rev. {\bf D59} (1999) 115004.
\bibitem{ckp}D. Chang, W. Keung, A. Pilaftsis, Phys.Rev.Lett. {\bf
    82} (1999) 900.; Erratum-ibid.{\bf 83} (1999) 3972; D. Chang,
  W. Chang and W. Keung, hep-ph/9910465.
\bibitem{pil} A. Pilaftsis, Phys. Rev. D {\bf 58}  (1998) 096010;
  A. Pilaftsis, Phys.  Lett. {\bf 435B} (1998) 88; D. A. Demir, hep-ph/9901389; D. A. Demir,
  hep-ph/9905571.
\bibitem{bekl}M. Brhlik, L. Everett, G.L. Kane and J. Lykken,
  Phys.Rev.Lett. {\bf 83} (1999) 2124; hep-ph/9908326.
\bibitem{hg} H. Goldberg, Phys. Rev. Lett. {\bf 50} (1983) 1419.
\bibitem{ks99}S. Khalil and Q. Shafi, hep-ph/9904448.
\bibitem{heavy}S.~Dimopoulos and G.F.~Giudice, Phys.~Lett.~{\bf B357}
(1995) 573 ; A.~Cohen, D.B.~Kaplan and A.E.~Nelson,  Phys.~Lett.~{\bf B388}
(1996) 599 ; A.~Pomarol and D.~Tommasini, Nucl.~Phys.~{\bf B488} (1996)  3.
\bibitem{pw99}A. Pilaftsis and C.E.M. Wagner,  Nucl.Phys. {\bf B553}
  (1999) 3.
\bibitem{ffo}
T.\ Falk, A.\ Ferstl, and K.A.\ Olive, Phys. Rev. {\bf D59} (1999)
055009; T.\ Falk, A.\ Ferstl, and K.A.\ Olive, hep-ph/9908311.
\bibitem{cin}
U.\ Chattopadhyay, T.\ Ibrahim, and P.\ Nath,  Phys. Rev. {\bf D60}
(1999) 063505;
P. Gondolo and K. Freese, hep-ph/9908390; S.Y. Choi, hep-ph/9908397.
\end{thebibliography}
\end{document}